\documentclass{sigchi}
\pdfoutput=1

\CopyrightYear{2016}
\setcopyright{acmlicensed}
\doi{http://dx.doi.org/10.475/123_4}
\isbn{123-4567-24-567/08/06}
\conferenceinfo{L@S'17,}{Apr 20--21, 2017, Cambridge, MA, USA}
\acmPrice{\$15.00}

\usepackage{balance}       
\usepackage{graphics}      
\usepackage[T1]{fontenc}   
\usepackage{txfonts}
\usepackage{epstopdf}
\usepackage{mathptmx}
\usepackage{color}
\usepackage{booktabs}
\usepackage{textcomp}

\usepackage{graphicx}
\usepackage{comment}
\usepackage{amsmath,xcolor}
\usepackage{graphicx}
\usepackage{enumitem}
\usepackage{pifont}
\usepackage{xspace}
\usepackage{url}
\usepackage{adjustbox}
\usepackage{array}
\usepackage{multirow}
\usepackage[T1]{fontenc}
\usepackage{booktabs}
\usepackage{lipsum}
\usepackage{transparent}
\usepackage{siunitx}
\usepackage{textcomp}

\usepackage{times,listings,amsfonts}

\usepackage{microtype}        
\usepackage{ccicons}          
	
\def\EmbedGrader{EmbedInsight\xspace} 

\usepackage{todonotes}

\def\plaintitle{SIGCHI Conference Proceedings Format}

\def\emptyauthor{}
\def\plainkeywords{Authors' choice; of terms; separated; by
  semicolons; include commas, within terms only; required.}

\makeatletter
\def\url@leostyle{%
  \@ifundefined{selectfont}{
    \def\UrlFont{\sf}
  }{
    \def\UrlFont{\small\bf\ttfamily}
  }}
\makeatother
\def\pprw{8.5in}
\def\pprh{11in}

\definecolor{linkColor}{RGB}{6,125,233}

\begin{document}

\title{\EmbedGrader: Automated Grading of Embedded Systems Assignments}

\numberofauthors{1}
\author{
  \alignauthor{Hao Li$^1$, Bo-Jhang Ho$^2$, Bharathan Balaji$^2$, Yue Xin$^2$, Paul Martin$^2$, Mani Srivastava$^2$\\
  \vspace{2mm}
    \affaddr{$^1$Zhejiang University \hspace{8mm} $^2$University of California, Los Angeles}\\
    \email{$^1$elvislee@zju.edu.cn, $^2$\{bojhang,bbalaji,yuexin,pdmartin,mbs\}@ucla.edu}}\\
}


\maketitle

\begin{abstract}
Grading in embedded systems courses typically requires a face-to-face appointment between the student and the instructor because of experimental setups that are only available in laboratory facilities. Such a manual grading process is an impediment to both students and instructors. Students have to wait for several days to get feedback, and instructors may spend valuable time evaluating trivial aspects of the assignment. As seen with software courses, an automated grading system can significantly improve the insights available to the instructor and encourage students to learn quickly with iterative testing. We have designed and implemented EmbedInsight, an automated grading system for embedded system courses that accommodates a wide variety of experimental setups and is scalable to MOOC-style courses. EmbedInsight employs a modular web services design that separates the user interface and the experimental setup  that evaluates student assignments. We deployed and evaluated EmbedInsight for our university embedded systems course. We show that our system scales well to a large number of submissions, and students  are satisfied with their overall experience.

\end{abstract}



\section{Introduction} 
Student assignments form an integral part of modern education. Formative feedback from assignments helps students to learn from their mistakes and informs instructors about common misconceptions~\cite{shute2008focus}. Automated assessment tools have therefore been a cornerstone of Learning@Scale~\cite{geigle2016exploration,lan2015mathematical} and education research~\cite{keuning2016towards}. Tools such as automated graders (autograders) can relieve instructors from time consuming manual grading and improve student learning~\cite{wilcox2015role}. Autograders have become commonplace for programming courses, with several commercial offerings~\cite{vocareum,gradescope,turingscraft}, and have about 65\% adoption according to a recent survey at SIGCSE~\cite{wilcox2016testing}. 

Automated assessment research has primarily focused on programming assignments~\cite{edwards2008web, singh2013automated}, mathematical equations~\cite{lan2015mathematical}, textual analysis~\cite{wang2008assessing} and simulation tools~\cite{juniwal2014cpsgrader}. We focus on courses that teach embedded systems concepts such as those in electronics, sensor networks, and real-time systems. It is common in these courses to evaluate student performance using hardware testbeds that measures the pertinent signals with instruments such as oscilloscopes. Instructors need to not only spend a significant amount of time preparing their experimental setup, but also need to spend time with each student to assess their performance. Valuable instructor time would be saved if an automated system could alleviate the repeated manual effort, and students can get immediate feedback on their performance instead of waiting on the instructor. The instructor could use the time saved to instead focus on common mistakes that students make and help them learn better.

Designing a general purpose autograder that works for most embedded systems courses is challenging as a wide variety of platforms and instruments are used by instructors. In addition, such a system would need to support features that have become standard in software autograders: (1) Students should get informative feedback. (2) Students should not be able to cheat or break the system. (3) The autograder should scale well for large classes. (4) The instructor should have the flexibility to design their own assignments. Several autograders have been proposed for embedded systems~\cite{juniwal2014cpsgrader,legourski2005system,netkow2010xest}, but these are either simulation-based~\cite{juniwal2014cpsgrader} or limited to specific hardware setups~\cite{valvano2016mooc,netkow2010xest}. 

We present \emph{\EmbedGrader}, a general purpose web services based modular and scalable autograder for embedded systems courses. \EmbedGrader consists of a primary server through which instructors specify their experimental setups and assignment details. The primary server communicates with a bank of hardware testbeds that are set up as per instructor requirements. Students submit their assignments through the primary server, which schedules execution on the testbeds, performs grading and returns the results back to the students. The modular design of testbeds enables us to support a diverse set of hardware platforms and measurement instruments. 

We have implemented and deployed \EmbedGrader for our university's graduate level embedded systems course. We show that \EmbedGrader can support multiple testbeds concurrently and scales well to bursts of hundreds of student submissions. Usage data and feedback from the course students showed that they successfully used the test cases available in \EmbedGrader to iteratively improve their programs. Students reported an overall experience score 4.0 out of 5 in our exit survey.

\section{Background and Motivation}
Embedded systems tackle the computing needs of special purpose devices such as self driving cars or heart pacemakers. The concepts of embedded systems are a critical part of electrical engineering and computer science education as they teach students the basics of cyber-physical systems, internet of things, real-time systems, wireless sensor networks, energy efficient computing, and firmware developments. At the heart of an embedded system is a logic device such as a microcontroller or an FPGA that communicates with peripheral devices or other embedded systems. Hence, typical course assignments ask students to write a microcontroller program which accomplishes tasks such as processing a sensor event to generate control signals, scheduling tasks given a time limit, or energy efficient communication in sensor networks. The performance of these student submissions are usually evaluated through high-fidelity equipment such as oscilloscopes, signal generators, power meters, and spectrometers.

The benefits of autograders have been well documented in literature~\cite{ala2005survey,keuning2016towards}. Here we highlight the challenges in grading embedded systems assignments and how autograders can alleviate them. From our experience\footnote{One of the authors has been teaching embedded systems courses for over 15 years, another author has designed embedded systems assignments for over 5 years.}, a typical grading process requires that the course lecturer or teaching assistant (we refer to both of them as ``instructors'') assembles an assignment specific experimental setup and allocates time slots to each student for assessment. This process naturally incurs extensive time investment from the instructor and only provides limited one-time feedback to the students. In addition, the assessment itself is challenging. As the instructor cannot anticipate all the student's mistakes, even a well designed rubric needs to be revised based on the overall performance for fair assignment of scores. Further, it is not possible for the instructor to perform additional testing retroactively.

The assignments themselves are designed with manual grading in mind. As an example, one of our past assignments requires microsecond level time synchronization between two devices, a fundamental concept in real-time distributed systems~\cite{sha1990priority}. To assess the synchronization performance, the two devices send periodic signals, and an oscilloscope is used to observe if their signals occur at the same time. Ideally, the instructor prefers to observe any drift between the signals for longer than a 15-minute time slot, and would like to test robustness by introducing external events or test across different types of clock sources. However, the cumbersome experimental setups and lack of time limit such analyses. 

Hence, an embedded systems autograder has the potential to not only improve feedback loops and save instructor time~\cite{wilcox2015role}, but also open up opportunities for systematic analysis, new types of experimentation, retroactive testing, precise feedback, and fair assessment. 




\section{Related Work}
Automated grading has been a topic of research since 1960~\cite{hollingsworth1960automatic}, and there exists a rich literature of autograders for software~\cite{edwards2008web,singh2013automated}, text analysis~\cite{larkey1998automatic,wang2008assessing}, mathematics~\cite{sangwin2007assessing,lan2015mathematical} and pronunciation~\cite{franco1997automatic}. Many of the autograding concepts have been integrated into commercial services like Vocareum~\cite{vocareum}, Gradescope~\cite{gradescope}, and are part of MOOCs like Coursera~\cite{coursera} and edX~\cite{edx}. However, these systems provide little to no support for hands-on hardware based embedded systems courses.  

Some of the recent embedded systems MOOCs have proposed simulation based autograders. CPSGrader~\cite{juniwal2014cpsgrader} creates a detailed virtual world of a robot rover and its environment, and students' programs are analyzed for the accuracy of control of the robot. The simulation is detailed enough to capture realistic effects, and students' programs could be applied to real robots with minor modifications. Another embedded systems MOOC~\cite{coursera_cooja} uses the Cooja sensor network simulator~\cite{osterlind2006cross} for its assignments. 
Such simulation based autograders naturally provide complete control over the assignment, expose unique learning opportunities and can be scaled like software autograders. However, students in such courses do not get a hands-on hardware experience.

It is difficult to completely replace hands-on experience with simulators. Creating scalable simulators which can accurately capture complex real world phenomena takes significant effort. For example, modeling all aspects of a real hardware such as signal jitter, clock drift, and button debounce is challenging. As another example, the Cooja simulator models network behavior in the Contiki operating system~\cite{dunkels2004contiki}, but it would take extensive engineering effort to integrate aspects such as wireless radio propagation, or variations in hardware performance. Even in embedded systems industry, simulations are used for the design stages, but real hardware testing is conducted before a product is released. Therefore, it is imperative that students learn to comprehend the limitations of simulations and work with real hardware.

Valvano et al. have created embedded systems MOOCs which provide hands-on hardware assignments~\cite{utaustin_embedded_sys,valvano2016mooc}. They worked with industrial manufacturers and vendors to create custom hardware and software tools for MOOC scale. They provide specialized compiler tools with custom dynamic-link libraries that instrument student programs to provide both simulation and  hardware based debugging support. Students need to set up the hardware on their own and manually provide inputs such as pressing buttons as per given instructions. An external library generates a score which is submitted to MOOC website by students. The grading process relies on student honesty and does not prevent cheating. In their system, the extensive instrumentation of the student programs is required for each assignment.
In \EmbedGrader, the assessment is performed in a controlled environment, and can support detailed instrumentation and prevent cheating. Our modular black box approach allows us to support different types of hardware and software tools without modifying student programs. As Valvano et al. take a distributed approach, the hardware needs to be inexpensive and easy to set up. With \EmbedGrader, however, the hardware needs to be set up just once and the instructor can use expensive instruments as students do not need to buy them.

Similar to \EmbedGrader, Legourski et al.~\cite{legourski2005system} created an embedded systems autograder that treats the student programs as a black box, and supplies inputs and captures outputs for assessment. However, their system only supports a specific hardware setup for input/output, and the instructors need to learn a special C-like language for grading specification. In contrast, \EmbedGrader can support any IO device that communicates with a computer, and the grading scripts are language independent. Like \EmbedGrader, Xest~\cite{netkow2010xest} provides a web-based embedded systems autograder, but instructors have to use their customized operating system and instrument student code for testing.
To the best of our knowledge, all of the prior embedded systems autograders are either simulation based or hardware platform/operating system specific. \EmbedGrader is the first attempt to create a general-purpose embedded systems autograder that scales well to large student submissions.

\section{EmbedInsight Overview}
Our objective is to design a general purpose embedded systems autograder. We adhere to following design principles to achieve our goal:
\begin{itemize}[noitemsep,nolistsep]
\item \textbf{Automation:} We seek to make the entire grading process automated, from hardware setup to feedback to students and instructors. 
\item \textbf{Modularity:} We consider modularity to be critical for a \emph{general purpose} autograder that supports the ever evolving, diverse assortment of embedded systems platforms and instruments.
\item \textbf{Security and Reliability:} We need protection mechanisms which prevent students from breaking the system or snooping into other students' files. 
\item \textbf{Scalability:} The system should handle heavy load during deadlines and scale linearly to support MOOCs. 
\item \textbf{Formative feedback:} Feedback should support student learning and empower instructors to improve teaching~\cite{shute2008focus}.
\item \textbf{Fairness:} The assessment should be fair across students.  
\end{itemize}
In addition, our design is inspired by commercial autograder services such as Vocareum~\cite{vocareum} and Gradescope~\cite{gradescope}.

\subsection{System Architecture}
\label{sec:systemArchitecture}
\begin{figure}[t!]
\centering
  \includegraphics[width=1\columnwidth, bb=0 0 352 313]{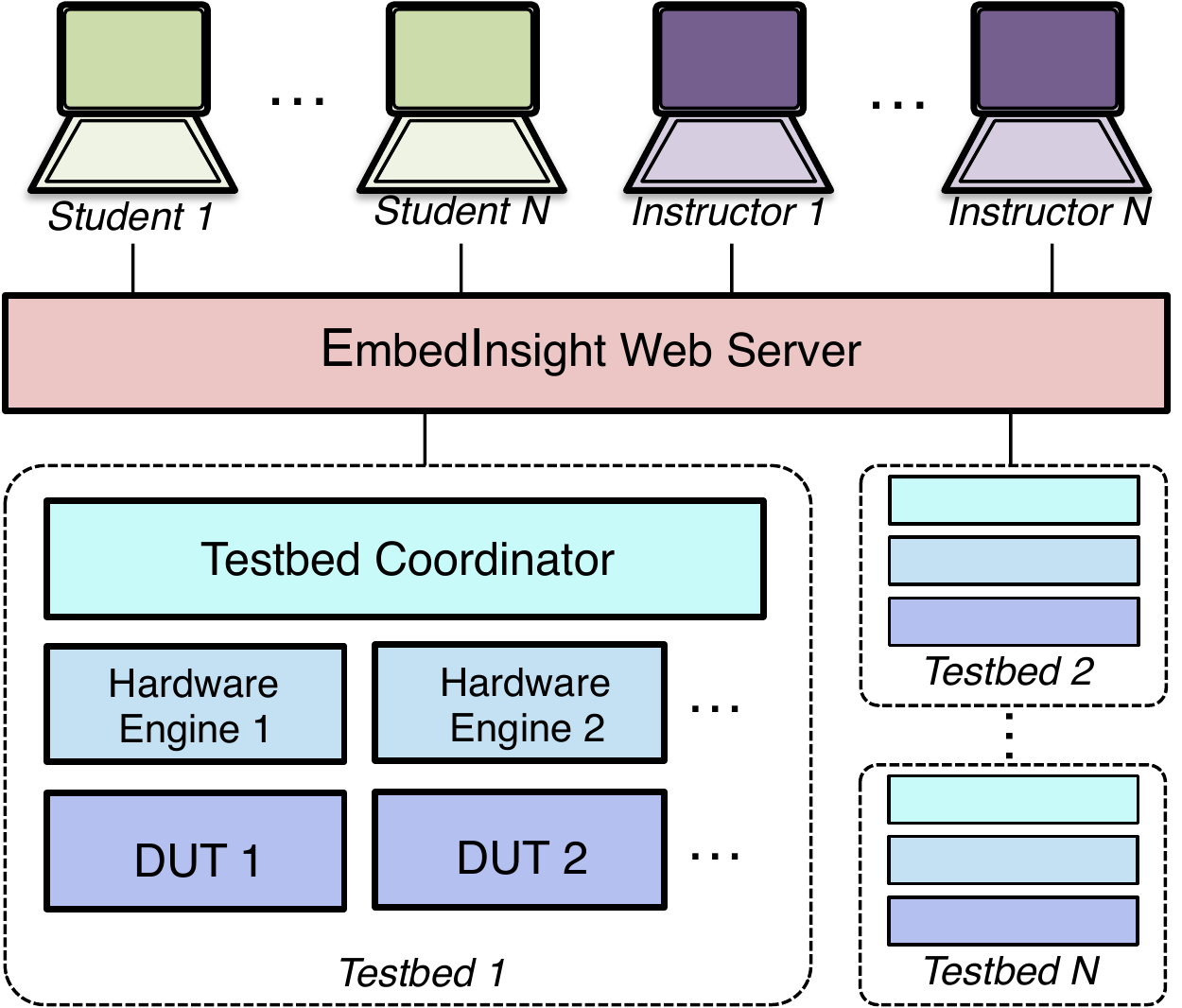}
  \caption{\EmbedGrader system architecture. The EmbedInsight Web Server acts as the primary server, and communicates with a network of testbeds to automate embedded systems grading at scale.}
  \label{fig:overview}
\end{figure}

We have designed \EmbedGrader as a composition of RESTful web services as it provides modularity, scalability, and performance~\cite{fielding2000architectural,hamad2010performance}. Figure \ref{fig:overview} shows the overview of our system. We have divided our system into two major components: (i) a \emph{primary web server} that serves as the UI and provides services similar to a software autograder, and (ii) a bank of \emph{testbeds} that replace the traditional experimental setups and performs assessment at scale. Each testbed itself is a web service and communicates with the primary server through RESTful APIs. The separation of hardware components into modular testbeds allows us to support different kinds of hardware arrangements at the same time.


Each testbed consists of three parts: (i) a testbed coordinator, (ii) hardware engine(s), and (iii) device(s) under test (DUT(s)). \emph{DUT} is the platform which executes student submitted programs. The testbed treats the DUT as a black box, and only provides inputs and captures outputs. 

The \emph{hardware engine} consists of all the components necessary to provide inputs to the DUTs and measure output signals. The hardware engine can consist of a range of components: microcontrollers, cameras, oscilloscopes, signal generators, spectrometers, power meters, etc. To support automation, we restrict the set of components to those that can communicate with an external computer and can be programmatically controlled. This is not a severe restriction as many commercial devices already support USB communication and have published APIs (e.g., ~\cite{SaleaeLogic}). We anticipate the list will continue to grow.

The \emph{testbed coordinator} serves as a bridge between the primary server and the components within the testbed. It hosts a web service to communicate with the primary server and contains device specific drivers to communicate with DUTs and hardware engines. The coordinator can be hosted in a small computer such as Intel NUC~\cite{NUC} or be implemented as an embedded web server~\cite{riihijarvi2001providing} to reduce costs. The coordinator manages all the resources within the testbed, and sends periodic status messages to the primary server. 

\subsection{PWM: An Example Assignment} 

\begin{figure}[t!]
\centering
  \includegraphics[width=1\columnwidth,trim={0.5cm 0 0.5cm 0},clip, bb=0 0 350 150]{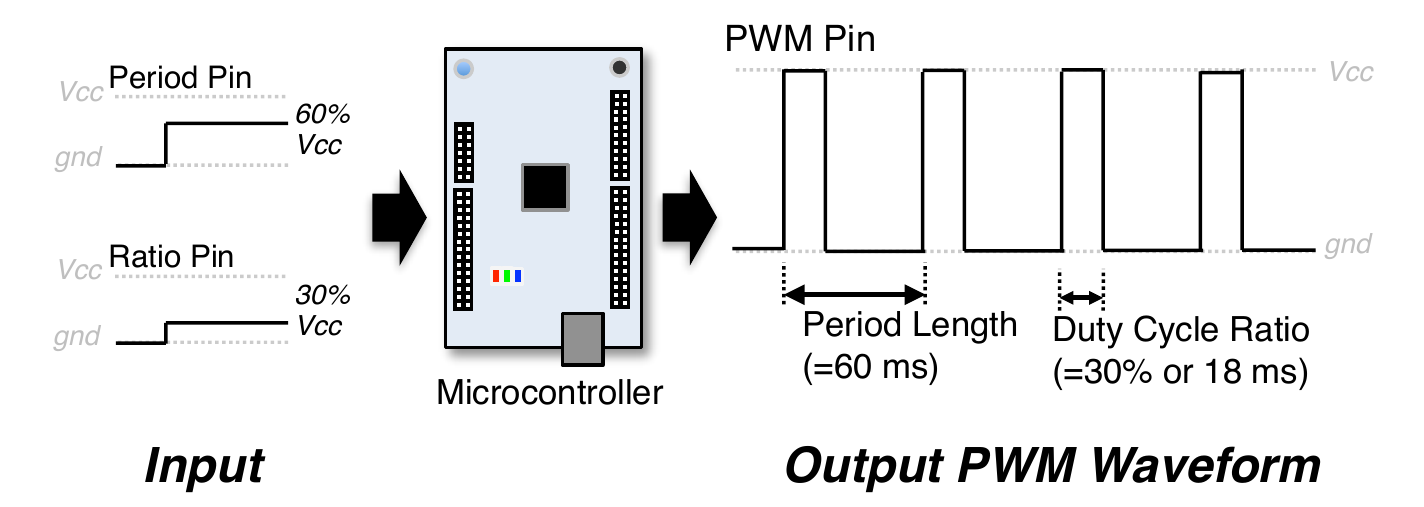}
  \caption{Depiction of the example PWM assignment. The microcontroller received the period and duty cycle ratio as input and generates the PWM waveform as output.}
  \label{fig:pwm}
\end{figure}

To concretely illustrate how \EmbedGrader works in practice, we consider Pulse Width Modulation (PWM) as a simple example. A PWM is a periodic digital waveform that modulates the duration for which the pulse remains high. A PWM signal is parameterized by: (i) \textbf{period}: the duration of a cycle, and (ii) \textbf{duty cycle ratio}: the proportion of time for which the pulse is high.  Figure \ref{fig:pwm} shows an example PWM waveform. PWM is used across many embedded systems applications to represent an analog value with a digital signal: control of motors, dimming of LED lights, etc~\cite{barr2001pulse}. 

In our assignment, we ask the students to generate a PWM waveform with a microcontroller given the period and duty cycle ratio as inputs. With a traditional grading system, the instructor would compile the student program on to the microcontroller and observe the PWM signal with an oscilloscope. She would verify the accuracy of the PWM by manually wiring different inputs using a breadboard.


\subsection{A Typical Workflow}
To release the PWM assignment using \EmbedGrader, instructors need to perform a set of initial steps. To start, an instructor registers with \EmbedGrader, creates a course and provides the assignment problem statement. When creating the assignment, an instructor needs to choose from a preregistered set of testbeds or register a new one. In the later case, the instructor has to precisely define the specifications of the hardware engine, the DUT, and how they are wired together. Once a new testbed is registered, a request is sent to the administrators, who then manually set up the testbed and inform the instructor when it is ready.

For each assignment, an instructor can generate several test cases, each with its set of inputs, outputs, and a grading script. For PWM, the instructor would specify the period and duty cycle ratio as input and ask to capture the PWM signal from the DUT as output. The grading script is an executable that generates a score and student feedback from the given input and output signals. The instructor can choose to make a subset of test cases public, similar to programming autograders~\cite{vocareum}. 

After the assignment is released, the students can view the problem statement, the specifications of the DUT and the public test cases. When students submit their programs, our system executes them on an available testbed and grades them using instructor provided scripts. After \EmbedGrader finishes grading, students can access the score, feedback and outputs captured for the public test cases. Students are allowed multiple submissions until the deadline so they can refine their programs based on the feedback from the test cases. \EmbedGrader also provides an overview of students performance to the instructors so they can monitor the progress, address issues, add test cases or extend the deadline.

\section{Design and Implementation}
We continue to use the PWM assignment as an example to explain the details of each as aspect of \EmbedGrader.

\subsection{Testbed}
\subsubsection{Device Under Test}

\begin{figure}[t!]
\centering
  \includegraphics[width=1\columnwidth, bb=0 0 430 250]{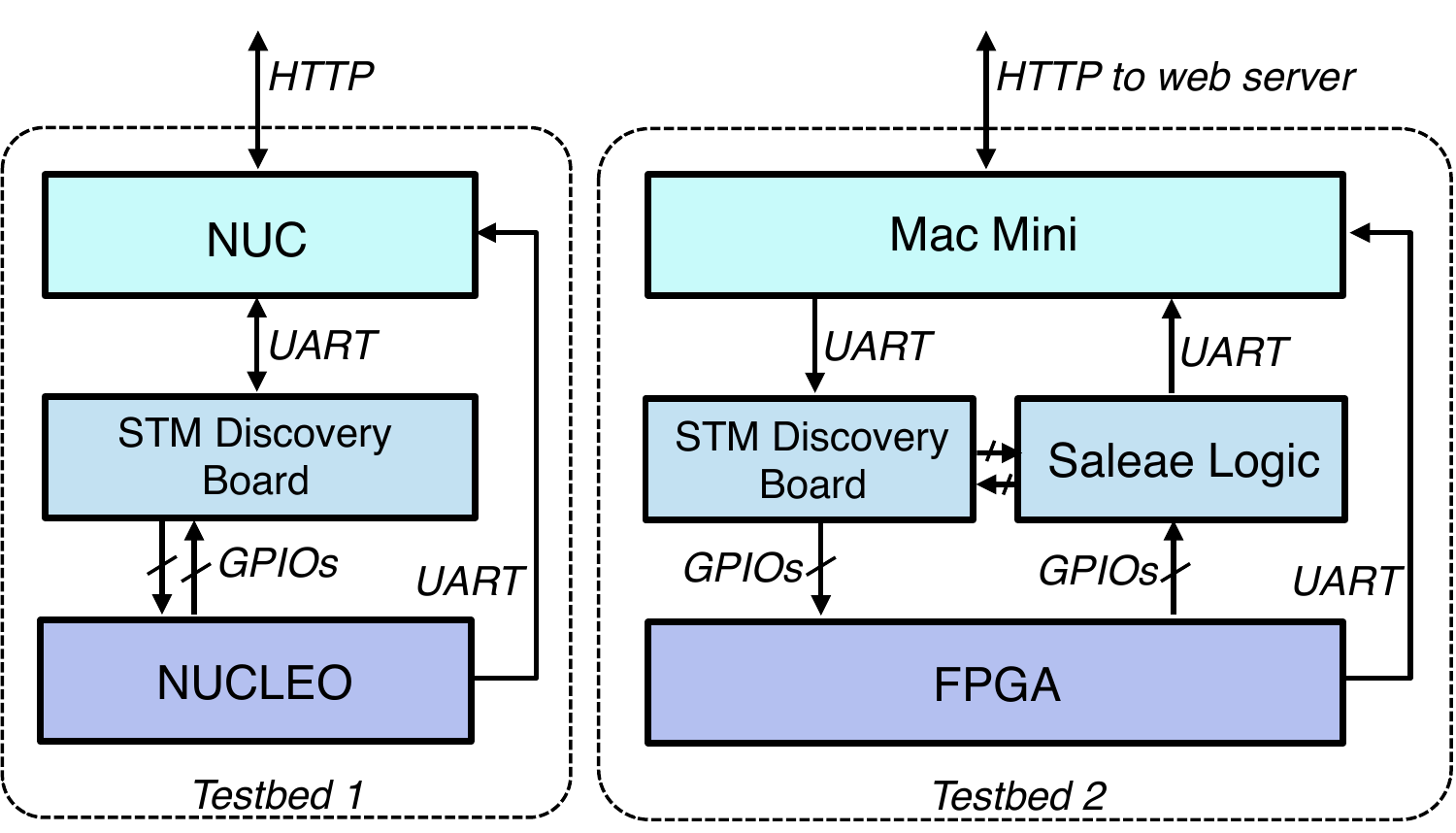}
  \caption{Experimental testbeds. Testbed 1 is used for the PWM generator assignment. Testbed 2 is an example of an alternative hardware setup within \EmbedGrader.}
  \label{fig:example_testbeds}
\end{figure}

Students programs have complete control over the DUT, and \EmbedGrader interfaces with it only through inputs/outputs. We made this design decision to support different DUTs without the overhead of instrumenting the programs. If the internals of student programs need to be analyzed, then existing techniques for software autograders~\cite{edwards2008web,singh2013automated,keuning2016towards} can be exploited and can be hosted as a module in the \EmbedGrader primary server. 

For the PWM assignment, we use the NUCLEO-F746GZ development board~\cite{NUCLEO} as our testbed. The PWM period and duty cycle ratio are provided using GPIO (General Purpose Input Output) pins. The inputs can change over time, and the program needs to change its output PWM accordingly. The quality of the program is judged based on how closely output PWM matches the given inputs and if it reacts quickly to input changes.

\subsubsection{Hardware Engine}
The hardware engine can consists of a set of devices that interfaces with the DUT inputs and outputs. Each of these devices has respective drivers to communicate with the testbed coordinator and can be used independent of each other. In real-time systems, an instructor may want to have fine-grained control over these devices. We propose a hardware engine hub that acts as a low-level controller for all devices. The hub can keep track of time in microseconds, synchronize multiple devices and even create a feedback loop by changing the input based on the DUT output. Such a setup enables a range of experimentation in networking, control systems, etc.  

We implement the hardware engine hub using STM32F407 Discovery board~\cite{STM32F407VG}, with an ARM Cortex M4 processor and 168MHz clock. We have designed this hub to be reusable across multiple assignments by establishing a communication protocol with the testbed coordinator and providing input/output capabilities for digital and analog signals. The hub includes additional functionalities like resetting the DUT, start/stop timer, and streaming data to and from the testbed coordinator. With our initial optimizations, we can sample DUT analog or digital signals at 100K samples/second. With these capabilities, the hardware engine can directly interface with the DUT or indirectly with sensors, wireless radio, etc. 

The Testbed 1 in Figure \ref{fig:example_testbeds} shows the hardware setup for the PWM assignment. Our hardware engine streams a timeseries of PWM period, duty cycle ratio inputs from the testbed coordinator, and supplies these inputs to DUT, i.e., the NUCLEO board. The hardware engine captures the output PWM signal from the DUT at 5K samples/second and streams the data back to the testbed coordinator for the duration of the experiment. As the student program can have mistakes or spurious behavior, a fixed sampling rate and fixed duration of capture allows us to limit any negative effects on the hardware engine. We have designed the hardware engine to be simple, so it can efficiently perform I/O functions. Higher level functionalities such as uploading programs to DUT, analyzing the output and detecting malfunctions are performed by the testbed controller and the primary server.   


\subsubsection{Testbed Coordinator}
\begin{figure}[t!]
\centering
  \includegraphics[width=1\columnwidth, bb=0 0 400 360]{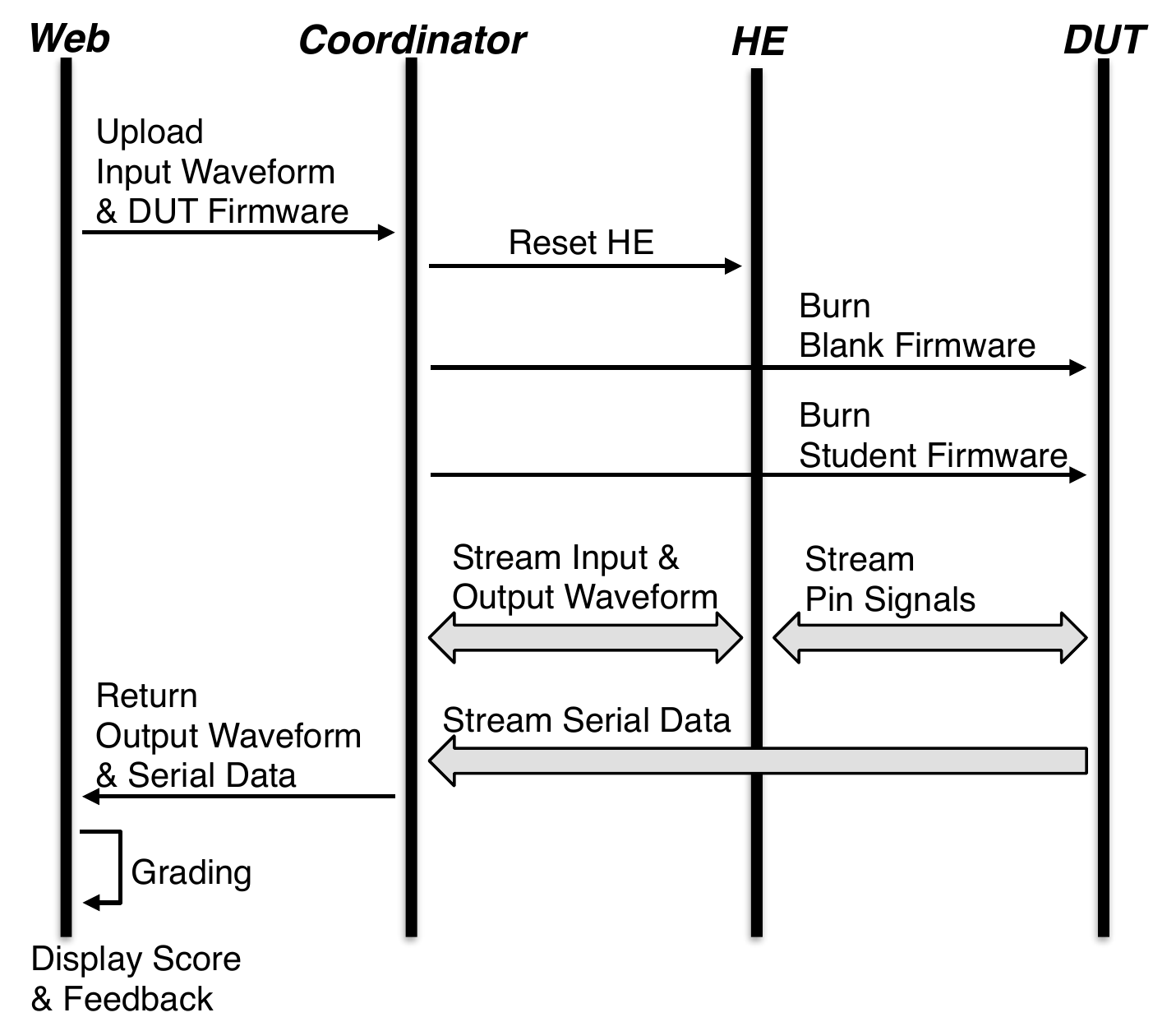}
  \caption{The flow of communication between the primary server (Web), testbed coordinator, hardware engine (HE) and device under test (DUT) for one session of grading the example PWM assignment.}
  \label{fig:grading_flow}
\end{figure}

The testbed coordinator presents the entire testbed as one cohesive unit to the primary server, and manages the hardware engine(s) and DUT(s). To facilitate communication with the primary server through platform independent APIs, we implement the testbed coordinator as a micro web service. The web service approach also provides the flexibility to host a diverse set of testbeds across geographically distributed labs, similar to remote access labs like iLab~\cite{hardison2008deploying}. We use SSL/TLS communication and access control to ensure that only verified testbeds can talk to the primary server.

When the hardware engines and the DUTs are being setup, we manually create a configuration file in the testbed controller that describes each component and how they are wired together. The configuration is advertised to the primary server periodically to indicate availability for testing and as a verification of the hardware specification. Although in our current implementation the wiring is manual, we can automate wiring in the future using multiplexers or FPGAs~\cite{legourski2005system}.  

Figure \ref{fig:grading_flow} shows the detailed flow of communication within the testbed for the PWM assignment. The testbed controller receives the command to assess a student program along with the firmware for the DUT and the input files that describes the PWM period and duty cycle ratio in a prespecified format. The coordinator resets the hardware engine (HE in Figure \ref{fig:grading_flow}), i.e., the STM32F407 board, to start each analysis from a clean state. We found that if the student program does not compile properly, the DUT firmware upload will fail and cause the NUCLEO board to continue executing its previously uploaded firmware. To prevent incorrect assessment, we first upload a blank firmware to the DUT and then upload the student program.  

The testbed coordinator initiates testing by streaming the input files to the hardware engines. The hardware engine interprets the input, sends appropriate digital signals to the DUT, captures the output PWM waveform and converts it to compact output format and streams it back to the testbed controller. In addition, the testbed controller connects directly to the DUT using USB (Figure \ref{fig:example_testbeds}) that captures the print statements in the program. The print logs are made available to students for debugging. We limit the rate of printing using UART protocol to 1Mbps, and as the duration of the experiment is fixed, the log file size is bounded. We have implemented the testbed controller on an Intel NUC~\cite{NUC} using the Klein web micro-framework~\cite{klein}.

\subsection{Primary Web Server}

\begin{figure}[t!]
\centering
  \includegraphics[width=1\columnwidth, bb=0 0 355 270]{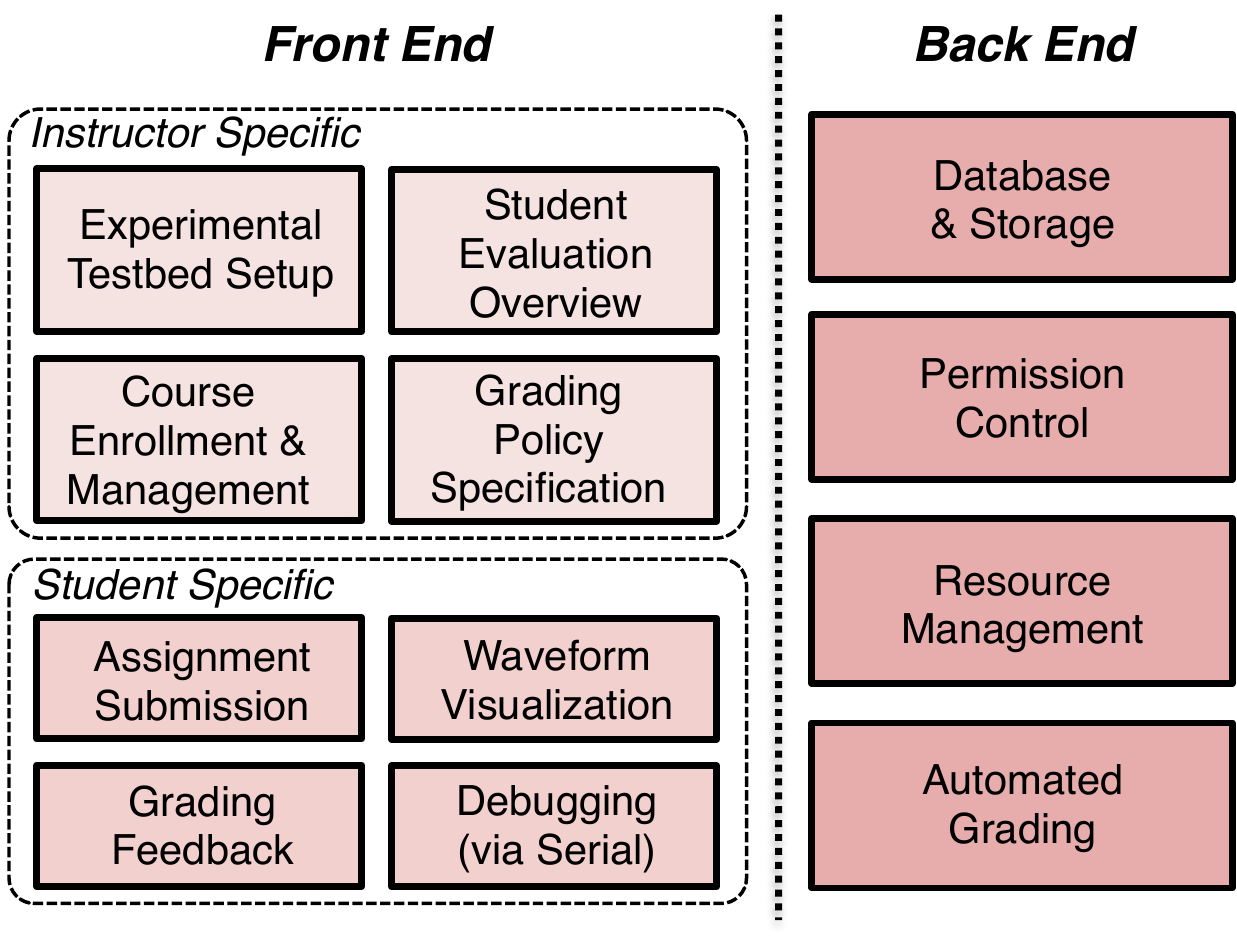}
  \caption{\EmbedGrader primary web server architecture. The frontend is responsible for providing interactive functionality to the instructor and students while the backend manages the access control and automated grading.}
  \label{fig:webserver}
\end{figure}

\begin{figure*}[t!]
\centering
  \includegraphics[width=1\linewidth, bb=0 0 720 250]{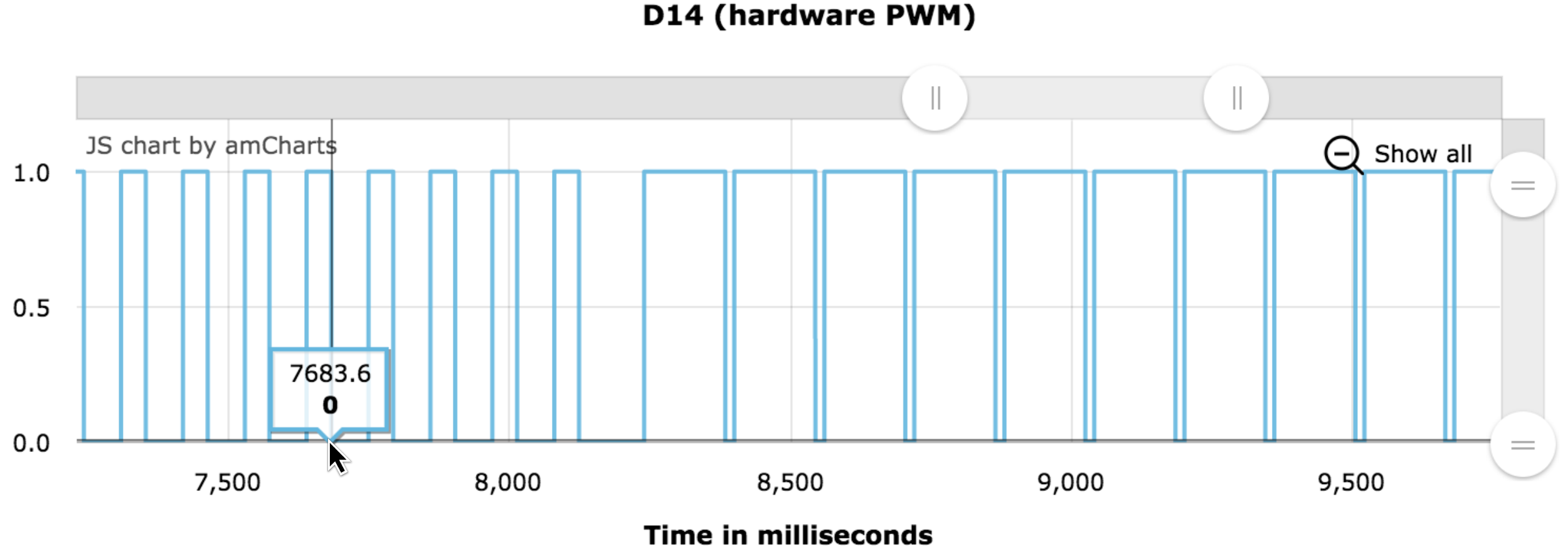}
  \caption{Screenshot of a PWM waveform generated from a student program as seen in \EmbedGrader user interface.}
  \label{fig:screenshot_waveform}
\end{figure*}

The primary server is the central point of communication for both instructors and students, and manages a network of testbeds. We have implemented the primary server using the Django web framework~\cite{django} and the Nginx web server. The front end of the web server handles the UI and the back end manages resources and exposes RESTful APIs to communicate with testbeds. Figure \ref{fig:webserver} shows an overview of the primary web server.

\subsubsection{Web Back End}
The back end is responsible for scheduling grading on a network of available testbeds, authentication and access control of testbeds and users, and information storage.  We have implemented the back end following the standard Model View Controller architecture~\cite{krasner1988description}, with PostgreSQL as our database. We assign user roles -- instructor and student -- for each course in the system, and carefully govern their permissions using role based access control~\cite{ferraiolo2003role}.


We implement the grading scheduler as an internal Django app that directly interfaces with the web server database and communicates with the testbeds using RESTful APIs. The scheduler continually polls for the pending student submissions that need to be graded, and allocates one thread to monitor and schedule work on each testbed in the system. As each grading task has to be completed without interruption, the scheduler can borrow strategies from non-preemptive scheduling literature~\cite{baruah2006non}. In our implementation, each thread polls for pending tasks with a randomized wait period to prevent collisions. Before starting the grading process in Figure \ref{fig:grading_flow}, the thread updates the database to mark the task as ``executing'' to avoid duplicate grading by other testbeds. The thread then coordinates with the testbed to obtain the output files after execution as explained earlier. The instructor given grading scripts are then applied on the output files to generate the score and feedback for the student.   



\subsubsection{Web Front End}

The web front end provides facilities the instructors to create courses, assignments, pick testbeds and specify their test cases. The list of available testbeds and their specification is populated based on periodic status messages sent by testbeds. The students can register and enroll in a course, view assignments, submit their program and view their grading results. The UI design is similar to commercial services~\cite{gradescope,vocareum}, and we implemented it using the Bootstrap framework~\cite{bootstrap}.  

Once an assignment is released, students are allowed multiple submissions until the deadline. After a submission is graded by the testbeds, for each test case \EmbedGrader provides multiple forms of feedback for students to analyze and revise their code. The instructor given script generates a score and a customized feedback that can point out errors in the output from DUT. Both the input and output files are made available along with visualizations of their waveforms. Figure \ref{fig:screenshot_waveform} shows a screenshot of the PWM waveform visualization. In addition, students can view the print logs that they can use for debugging. The instructor is provided with an overview of all the student submissions so they can monitor student progress. The instructor can choose to zoom in and observe the progress of individual students, analyze their outputs and give feedback. 

Moving forward, we plan to provide interactive debugging sessions to the students. We will provide a video streaming of the hardware testbed with an IP based camera to capture the LEDs, LCD displays. The students would reserve a testbed for a debugging session and create customized test cases to analyze their programs. Our current architecture allows for such interactive debugging, but we need to optimize the data flow between the DUT and the front end to provide a seamless responsive experience.

\section{Evaluation}
We demonstrate the performance of \EmbedGrader with a deployment in an embedded systems course and show how effectively the students used our system. We further conduct a series of microbenchmarks to evaluate the scalability and modularity of our system.

\subsection{Deployment in Our University Course}

\begin{figure}[t!]
\centering
  \includegraphics[width=1\columnwidth,trim={0.1cm 7cm 0 7.5cm},clip, bb=20 200 610 580]{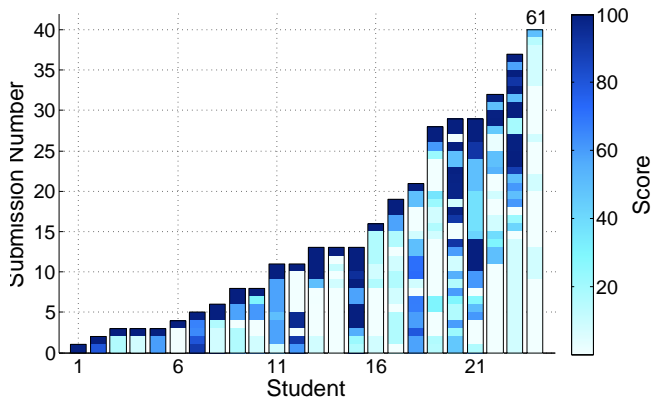}
  \caption{Score changes of each student across submissions. Hidden test cases are not included. The score change reflects how students iteratively modify their programs. The last student has 61 submissions.}
  \label{fig:progress}
\end{figure}

We deployed \EmbedGrader for our university's graduate level embedded systems course. There are 24 students in the class, and most of them  have an electrical engineering background. The instructor\footnote{The instructor is one of the paper authors} released the PWM assignment on \EmbedGrader, and had one testbed setup for grading as shown in Testbed 1 of Figure \ref{fig:example_testbeds}. The instructor provided one week to work on the homework, and accepted submissions to the system in the last four days. Due to development delays, we enabled the waveform visualization feature only one day before the deadline. The PWM assignment has three public, one semi-public, and two hidden test cases. The semi-public test cases show the final score like the public test cases but do not provide debugging information such as input/output waveforms and grading feedback. Students can view the details of the hidden test cases after the deadline is passed.

We define a \emph{session} as the duration for which the PWM period and duty cycle remains the same. Each test case consists of several sessions. Each session starts with supplying the PWM inputs, i.e., the period and the duty-cycle ratio, to the DUT. All sessions are graded independently and the scores are assigned based on the accuracy of the measured period and duty cycle ratio of the output PWM waveform. 

After the deadline, we asked the students to fill out a 2-minute survey to get feedback on \EmbedGrader. We asked about the usefulness of feedback, their positive/negative experiences and their overall experience. All the questions are either in a 5-point Likert scale or in free text format. 16 of the 24 students responded to the survey.

Figure \ref{fig:progress} shows an overview of the student submissions across 4 days. The color of the bar graph shows the score received by a student in each submission. The score only includes public and semi-public test cases, as hidden test scores are not visible to the students before the deadline. We received an average of 15.6 submissions per student and since PWM is a fairly simple assignment for graduate level, all the students achieve perfect scores in the visible test cases. However, only 4 students receive a score greater than 50 in their first submission. Our survey responses indicate that students at first misunderstood parts of the problem statement or how exactly their PWM waveform will be evaluated. The test cases helped them to iterate over their code, and helped them better understand the requirements. This is in stark contrast to the one-time grading policy in the previous versions of this course. With manual grading, the instructor needed to help students correct minor misunderstandings, had little time to provide conceptual feedback and fair assessment across students was difficult. Figure \ref{fig:progress} shows that \EmbedGrader can be a platform for students to test different approaches and understand their impact on the performance. Several students keep submitting codes even after getting full points. In the survey, students indicated that the test cases were useful for optimizing and refining their code.  
Finally, graphs like these when provided to the instructors provide a quick overview on the progress of the students, and helps identify students who may be struggling with the assignment. 

In our survey, several students appreciated that \EmbedGrader is responsive, robust and easy to use. Students found that the waveform visualization (Likert Scale: $3.6\pm1.45$) and the grading feedback ($4.18\pm0.91$) were especially useful for debugging. For the duration of the deployment, the average latency of grading a submission was 5.25 minutes and the median latency was 3.64 minutes. Some students were satisfied with the response time of \EmbedGrader ($3.37\pm1.58$), whereas others recommended that \EmbedGrader can generate preliminary results quickly instead of displaying complete grading results at the end of executions. Some students suggest the grading feedback could be more informative and would appreciate if \EmbedGrader can ``mark incorrect part of the waveform'' in the UI. 
When asked about their overall experience from 1-5 with 1 being ``Not Satisfied'' and 5 being ``Fully Satisfied'', none of the 16 respondents gave a \EmbedGrader a score of 1 or 2, and the average score was $4\pm0.89$.

\subsection{Scalability}

\begin{figure}[t!]
\centering
  \includegraphics[width=1\columnwidth, bb=0 0 360 220]{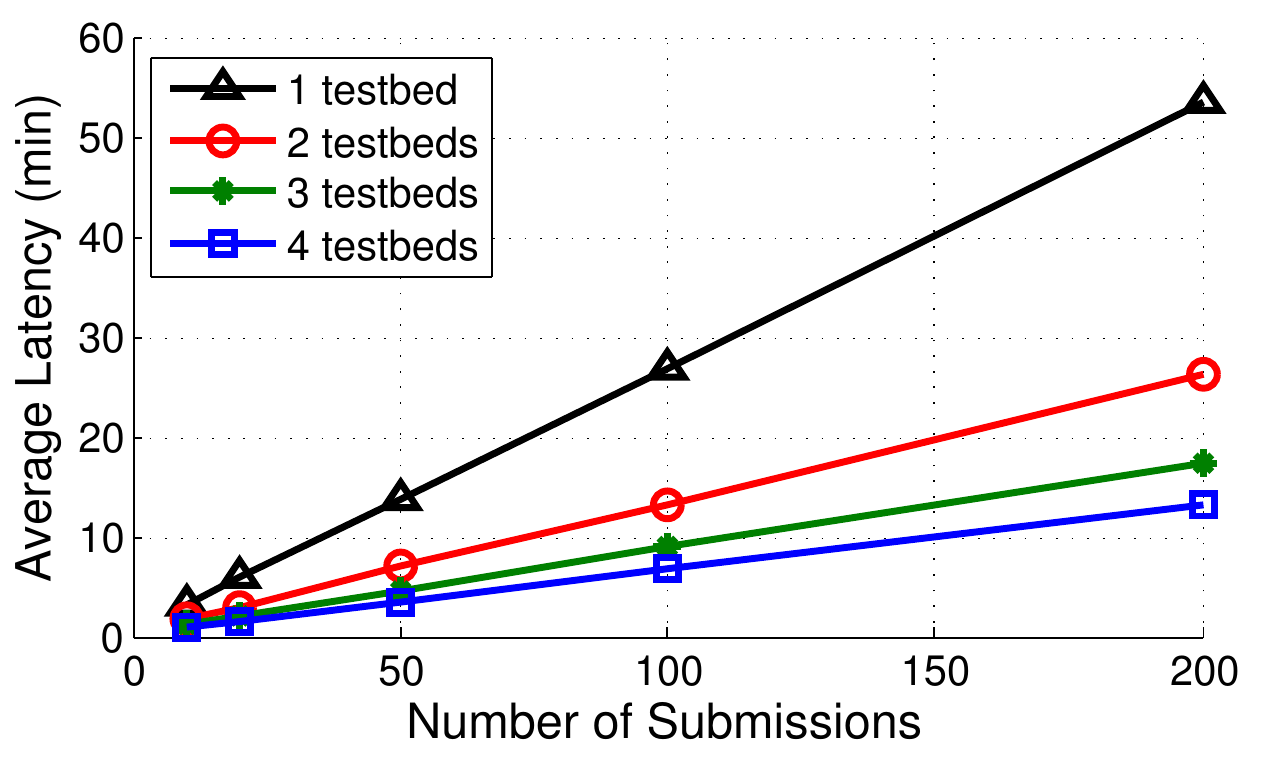}
  \caption{The average latency vs submission load under different number of testbeds. Latency is the user observed response time for grading a task. The latency increases linearly with increase in submissions, but addition of testbeds improves the latency quasi-linearly.}
  \label{fig:latency}
\end{figure}

\begin{figure}[t!]
\centering
  \includegraphics[width=1\columnwidth, bb=0 0 340 220]{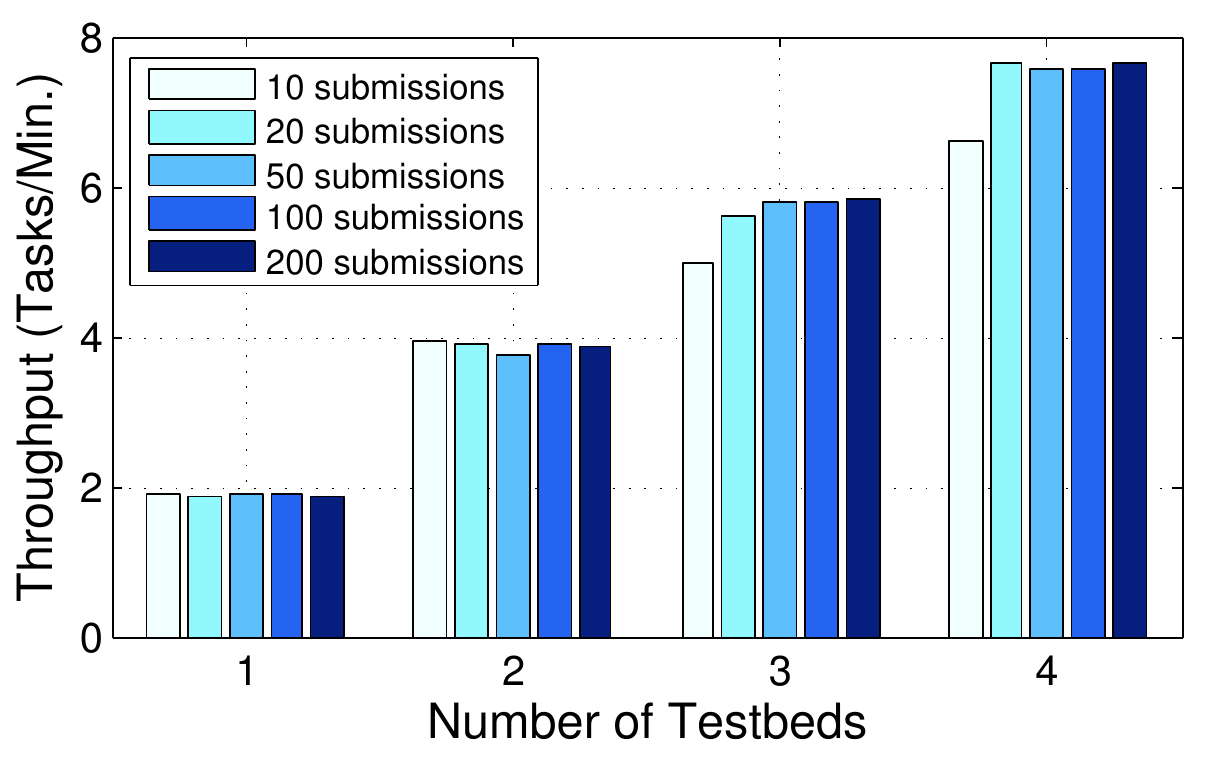}
  \caption{The throughput of \EmbedGrader with different number of testbeds with varying workloads. The workloads are determined by number of submissions that occur at the same time. The graph shows linear scaling in throughput with addition of testbeds.}
  \label{fig:throughput}
\end{figure}

To evaluate the scalability of \EmbedGrader with the addition of testbeds, we performed a microbenchmark to systematically measure the latency and throughput with varying load on the system. We created an isolated instance of \EmbedGrader that does not receive any external traffic, and created a PWM assignment with a single test case that analyzes PWM output for 10 seconds. An automated script submits programs to the \EmbedGrader, varying the number of submissions at a time from 10 to 200. We repeated this test with 1 to 4 testbeds.


Figure \ref{fig:latency} shows the average latency as we vary the workloads and number of testbeds. We define latency as the duration between the time of program submission and the time at which grades are available. The graph shows that the average latency scales linearly with submission load. Figure \ref{fig:throughput} shows the throughput of \EmbedGrader with varying load and testbeds. In general, the throughput is proportional to the number of testbeds. However, the throughput decreases when the number of submissions is small. This is because the number of submissions is not sufficient to saturate all the testbeds. When the number of submissions increases, \EmbedGrader delivers consistent throughput, which shows the stability of our system. Nevertheless, our system takes around 30 seconds on average to grade a 10-second waveform, indicating significant overhead. The major bottleneck happens to be in the testbed side since burning programs to DUTs take significant amount of time. In the future, we plan to provide interactive debugging sessions, and low latency will be a core requirement for a satisfying user experience. Therefore, latency caused by network communication, database, and grading task scheduling should be optimized to reduce overheads.


\subsection{Modularity}

\begin{figure}[t!]
\centering
  \includegraphics[width=1\columnwidth, bb=0 0 360 150]{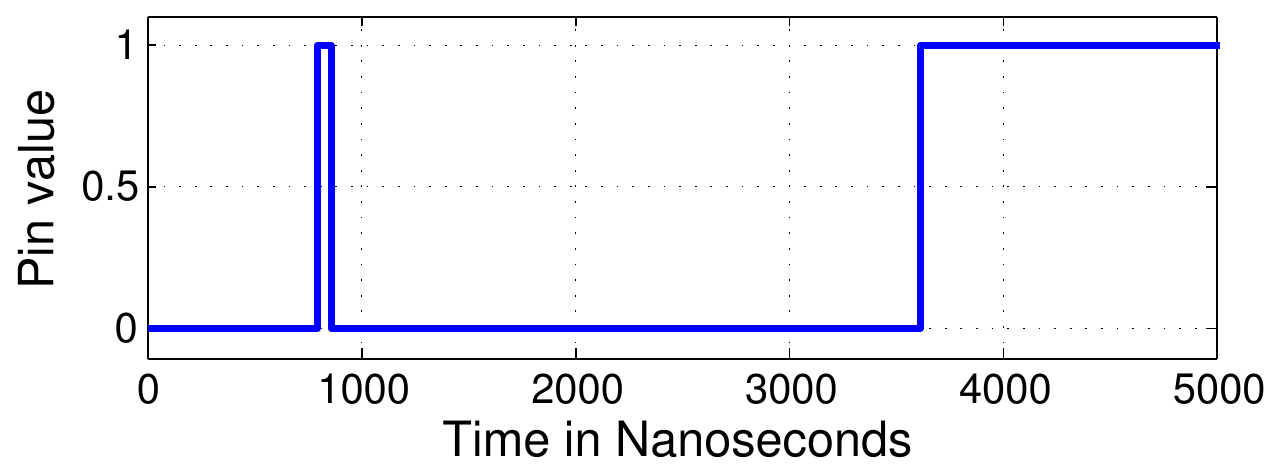}
  \caption{Signal jitters as observed in GPIO based PWM waveform when observed Saleae Logic scope. A spike spanning 62 nano seconds happens right before the pin is toggled.}
  \label{fig:jitter}
\end{figure}

We demonstrate the modularity of \EmbedGrader by showing that we can swap out multiple components in our testbeds, and still support the autograding functionalities. As an example, we replace our NUCLEO board with an FPGA as the DUT. We use the Lattice iCE40 FPGA\footnote{http://www.latticesemi.com/en/Products/FPGAandCPLD/iCE40.aspx}, which is programmable over USB with an open-source compilation procedure\footnote{IceStorm Project http://www.clifford.at/icestorm/}. We connect this FPGA to our testbed, program it with a command from the testbed controller, and capture its PWM waveform from the hardware engine.

We also show that adding components in hardware engine layer is possible to increase the accuracy of waveform measurements. To illustrate this, we use Saleae Logic \cite{SaleaeLogic}, a high-resolution signal analyzer that can sample 16M samples/second and capture waveform changes in the nanosecond level. At such a granular resolution, additional observations can be made about the DUT output signal. For example, to fulfill our PWM assignment requirement, students can use hardware supported PWM pins to generate the their output waveform, or programmatically toggle a GPIO pin with a hardware timer based interrupt. The hardware supported PWM produces a precise output signal whereas GPIOs, which are not designed for high-speed toggling, will have jitters, i.e. tiny spikes, in their output signal. We used the Saleae Logic oscilloscope to capture these jitters from a GPIO based PWM waveform. The jitter is visualized in Figure \ref{fig:jitter} -- the duration of the spike is 62 nanoseconds. With this augmented hardware engine setup, the instructor can tell whether students use software or hardware PWM.

\section{Conclusion and Future Work}
Grading in embedded systems courses is a slow, painful process that takes up significant amount of instructor time, but provides limited feedback to students. While autograders have advanced the learning experience in software and text based courses, hardware autograders have been limited to simulations or specific experimental setups. We have designed \EmbedGrader as a general purpose embedded systems autograder which can be reused across a range of different hardware configurations. \EmbedGrader decouples the user interactions and the hardware testbeds using web service composition, and scales in a modular manner to support arbitrary size classes. Our evaluation shows that \EmbedGrader scales well with addition of testbeds. Our deployment in a 24-student embedded systems course demonstrated that students successfully used \EmbedGrader to debug and optimize their programs, and were satisfied with their overall experience.

With \EmbedGrader, we have taken the first steps towards a modular, general purpose autograder for hardware based courses. While implementing and using the system, we identified several components that can be improved. Much of the embedded systems inputs/outputs are specified using analog and digital signals. Although there have been attempts to formalize waveform specification in literature~\cite{willink2001waveform}, there is little support to easily specify the characteristics of these signals with a programming language. In future work, we seek to create tools and schemata that will help instructors easily express a diverse set of waveforms. Similarly, it is difficult to specify digital/analog signal based grading requirements or generate feedback for the students with conventional tools. We seek to exploit techniques such as time automata~\cite{alur1994theory,larsen1997uppaal} from real-time systems community and integrate them into \EmbedGrader to ease assignment creation, so instructors can focus on learning aspects of an experiment rather than specification.

The field of embedded systems is just one of the many fields that uses hardware based assignments. Several other fields such as control systems, circuit design, signal processing, and telecommunication can benefit from autograders like \EmbedGrader. The equipment used in these areas need to be engineered so that they can be controlled using APIs, and then they can be modularized into remotely available testbeds~\cite{hardison2008deploying}. Furthermore, with Internet of Things, embedded systems components proliferate many aspects of our lives, from smart watches to smart meters. The \EmbedGrader architecture can be extended to support well instrumented smart environments to support experimentation in fields such as ubiquitous computing.


%
%
%
%
%


\bibliographystyle{SIGCHI-Reference-Format}
\bibliography{EmbedGrader}

\end{document}